# Doomed Worlds II: Reassessing Suggestions of Orbital Decay for TrES-5 b

Marvin Rothmeier ,[1] Elisabeth R. Adams ,[2] Karsten Schindler ,[1,3] André Beck ,[1] Brian Jackson ,[4]
Jeffrey P. Morgenthaler ,[2] Amanda A. Sickafoose ,[2] Malia Barker ,[4] Luigi Mancini ,[5,6,7]
John Southworth ,[8] Daniel Evans ,[8] and Alfred Krabbe [1]

[1]Institute of Space Systems, University of Stuttgart, Pfaffenwaldring 29, 70569 Stuttgart, Germany
[2]Planetary Science Institute, 1700 E. Ft. Lowell, Suite 106, Tucson, AZ 85719, USA
[3]SOFIA Science Center, NASA Ames Research Center, Moffett Field, CA 4035, USA
[4]Department of Physics, Boise State University, 1910 University Drive, Boise ID 83725-1570 USA
[5]Department of Physics, University of Rome "Tor Vergata", Via della Ricerca Scientifica 1, 00133 Rome, Italy
[6]INAF – Turin Astrophysical Observatory, Via Osservatorio 20, 10025 Pino Torinese, Italy
[7]Max Planck Institute for Astronomy, Königstuhl 17, 69117 Heidelberg, Germany
[8]Astrophysics Group, Keele University, Staffordshire, ST5 5BG, UK

## ABSTRACT

TrES-5 b is one of only three ultra-hot Jupiters (UHJs) with suggestions of a possibly decreasing orbital period that have persisted through multiple independent analyses (G. Maciejewski et al. 2021; S. R. Hagey et al. 2022; E. S. Ivshina & J. N. Winn 2022; W. Wang et al. 2024; L.-C. Yeh et al. 2024). While WASP-12 b's decreasing period is well-explained by tidally induced orbital decay (K. C. Patra et al. 2017), and stellar acceleration has been proposed for WASP-4 b (L. G. Bouma et al. 2020), the cause of the apparent trend for TrES-5 b has not been satisfactorily explained. This work extends the previous observations with 14 new ground-based transits from 2016-2024 and two newly-published midtimes for data from 2007 and 2009. Four TESS sectors (75, 77, 82, and 84) have also been included for the first time. With the new data, the case for a decreasing orbital period is much weaker than before. The revised rate of period change, $\dot{P} = -5.3 \pm 2.2$ ms yr$^{-1}$, is less than half that was found in previous work and the preference for a quadratic over a linear model, as measured through $\Delta\mathrm{BIC}_{\mathrm{LQ}}$, has been falling since 2020, with a current value of 11. Furthermore, these results are not robust to outliers; removing a single early transit midtime causes the effect to vanish ($\Delta\mathrm{BIC}_{\mathrm{LQ}} = -1$). Additionally, no significant periodic signals in the transit timing data are identified. The current data are well explained by a linear ephemeris.

## 1. INTRODUCTION

As observational baselines have passed the decade mark for many systems, a growing number of ultra-hot Jupiters around main-sequence stars have been presented with suggestions of potentially decreasing orbital periods. The Short Period Planets Group (SuPerPiG) has been systematically reanalyzing existing data and taking new observations to search for signs of non-Keplerian orbital evolution. Of 43 planets examined in the previous paper of the "Doomed Worlds" series (E. R. Adams et al. 2024), only two had statistically significant decreasing orbital periods: (1) WASP-12 b (first noted to be decaying by G. Maciejewski et al. 2016; K. C. Patra et al. 2017), and (2) TrES-1 b (previously noted as possibly decreasing by S. R. Hagey et al. 2022; E. S. Ivshina & J. N. Winn 2022), for which the apparent rate of change observed is much larger than expected for orbital decay.

The present study is focused on a single target, namely TrES-5 b, a UHJ not included in E. R. Adams et al. (2024) since new observations had not yet been taken by the group. TrES-5 b orbits a main-sequence K dwarf ($R_\star = 0.9\ R_\odot$, $M_\star = 0.9\ M_\odot$, log g = 4.5, $V = 13.7$) and is one of the earlier known transiting exoplanets (G. Mandushev et al. 2011), but follow-up observations have been more limited compared to its contemporaries. Searches for possible transit timing variations (TTVs) can be used to look for companion planets or stars and to reveal long term changes caused

Corresponding author: Marvin Rothmeier
rothmeier.astro@gmail.com

by e.g. tidal decay. An early TTV analysis was published that found evidence for short-term periodic perturbations in E. N. Sokov et al. (2018), which future work has failed to replicate. A potentially decreasing orbital period was first noted by G. Maciejewski et al. (2021), and evidence has been seen in subsequent analyses drawing from largely the same data (S. R. Hagey et al. 2022; E. S. Ivshina & J. N. Winn 2022; W. Wang et al. 2024; L.-C. Yeh et al. 2024). In particular, G. Maciejewski et al. (2021) observed a deviation from a Keplerian orbit, which was fit equally well by either a long-term periodic model or a quadratic model with a decreasing orbital period. The best fit orbital decay rate of $\dot{P} = -20.4 \pm 4.5\,\mathrm{ms\,yr^{-1}}$ implied a tidal quality factor of $Q'_* = 1.4 \times 10^4$, much smaller than expected for typical dwarfs (e.g. A. S. Bonomo et al. 2017). Apsidal precession was ruled out using a joint radial velocity (RV) and TTV model. As discussed by D. Ragozzine & A. S. Wolf (2009), the rate of apsidal precession in UHJs is sensitive to the planet's internal structure, parameterized by the planetary Love number $k_2$. For TrES-5 b, the value of $k_2$ inferred under the precession hypothesis was found to be unphysically high. The most likely remaining explanation was an acceleration of the systemic barycenter along the line of sight. This could be caused by a massive, wide-orbiting companion, an architecture that may be common among hot Jupiter systems (L. G. Bouma et al. 2020); however, no evidence of such a companion has yet been detected.

The goal of this work is therefore twofold: (1) to re-examine the evidence in the literature that the period of TrES-5 b is in fact changing, by incorporating new transit observations and analyses; and (2) to inspect which of the three potential models – orbital decay, orbital precession, or line-of-sight acceleration – could best account for the observed deviations in the transit timing residuals, if any.

## 2. TRANSIT EPHEMERIDES

When the orbital period remains constant, the mid-transit time, or midtime, is a linear function of the number of elapsed orbits. The linear ephemeris can be described by

$$T_E = T_{0,\mathrm{lin}} + P_{\mathrm{lin}} \cdot E, \tag{1}$$

where $T_E$ is the midtime at epoch $E$, $T_0$ is the reference midtime at $E = 0$ for the given model, and $P_{\mathrm{lin}}$ is the constant orbital period.

If the orbital period decreases over time, transits occur slightly earlier than expected based on a Keplerian orbit, introducing a quadratic term. The quadratic ephemeris can be described by

$$T_E = T_{0,\mathrm{quad}} + P_{\mathrm{quad}} \cdot E + \frac{1}{2}\frac{\mathrm{d}P}{\mathrm{d}E}E^2, \tag{2}$$

where $\mathrm{d}P/\mathrm{d}E$ is the rate of change of the quadratic orbital period, $P_{\mathrm{quad}}$, per epoch. The period change per orbit may be converted into the period change per unit time (in $\mathrm{ms\,yr^{-1}}$) using

$$\dot{P} = \left(\frac{\mathrm{d}P}{\mathrm{d}E}\right) \cdot \left(\frac{365.25}{P_{\mathrm{quad}}}\right)(86400 \cdot 1000). \tag{3}$$

In the presence of a periodic variation in the orbit – potentially caused by apsidal precession or line-of-sight acceleration – a sinusoidal model can be fit to the transit timing data without specifying the physical mechanism. This model is given by

$$T_E = T_{0,\mathrm{sin}} + A\sin(f \cdot E + \delta), \tag{4}$$

where $T_{0,\mathrm{sin}}$ is an arbitrary time offset, $A$ denotes the amplitude, $f$ is the frequency, and $\delta$ represents the phase shift.

The best fit for each model is achieved using the standard Levenberg-Marquardt algorithm (K. Levenberg 1944; D. W. Marquardt 1963). To determine which model is preferred, the Bayesian Information Criterion (BIC; G. Schwarz 1978) can be used to compare different models following the approach of A. R. Liddle (2007). It can be calculated as

$$\mathrm{BIC} = k\ln N + \chi^2, \tag{5}$$

where $k$ is the number of model parameters (2 for the linear model, 3 for the quadratic model, and 4 for the sinusoidal model), $N$ is the number of data points, and $\chi^2$ quantifies the goodness of the fit as a calculated statistic representing the difference between the data and the model. A lower BIC indicates a model is preferred because it provides a better balance between model fit and complexity. To compare models X and Y, the difference in BIC is computed as

$$\Delta\mathrm{BIC_{XY}} = \mathrm{BIC_X} - \mathrm{BIC_Y}, \tag{6}$$

where a positive $\Delta\mathrm{BIC_{XY}}$ thus favors model Y. This work uses letters L, Q, and S to refer to linear, quadratic, and sinusoidal models, respectively. Per this definition, $\Delta\mathrm{BIC}$ values reported in most prior works correspond to $\Delta\mathrm{BIC_{LQ}}$ here.

An approximate estimate for the uncertainty of $\Delta\mathrm{BIC_{XY}}$ is given by

$$\sigma^2_{\Delta\mathrm{BIC_{XY}}} \approx 2\left(N - \mathrm{k}_X - \mathrm{k}_Y\right), \tag{7}$$

where the degrees of freedom (DOF) for each model are $\mathrm{DOF}_L = N - 2$, $\mathrm{DOF}_Q = N - 3$, and $\mathrm{DOF}_S = N - 4$. As discussed in B. Jackson et al. (2025), the variance for $\Delta\mathrm{BIC}$ depends on the degrees of freedom for the two models being compared, the covariance between the best-fit $\chi^2$ values for those two models, and terms representing the degree of mismatch between the actual behavior of the transit-timing data and the model used to fit the data. Equation 7 does not include these latter two terms but represents a reasonable minimum estimate for the variance. A $\Delta\mathrm{BIC}$ that exceeds zero by at least Equation 7 suggests an emerging non-Keplerian trend but is not, on its own, decisive. Additional analysis and observations would be required to confirm such a trend.

## 3. PRIOR ANALYSES

Previous studies on TrES-5 b have reported significant variability in the calculated values for $\dot{P}$ and $\Delta\mathrm{BIC_{LQ}}$, as illustrated in Table 1. All authors find a deviation from the linear ephemeris and suggest the potential presence of orbital decay in their analyzed time frame. However, many of these investigations (S. R. Hagey et al. 2022; E. S. Ivshina & J. N. Winn 2022; W. Wang et al. 2024; L.-C. Yeh et al. 2024) did not concentrate solely on TrES-5 b; instead, they employed semi-automated methods to analyze multiple exoplanetary systems. Data were drawn from a diverse collection including prior publications in the peer-reviewed literature, newly observed light curves, transits reported to the Exoplanet Transit Database (ETD; S. Poddaný et al. 2010), and photometry from TESS (G. R. Ricker et al. 2015). As highlighted in the first paper of this series (E. R. Adams et al. 2024), it is important to be cautious when compiling literature transit midtimes. Therefore, meticulously examining published midtimes and eliminating any inaccuracies was the first task of this work.

**Table 1.** Suggested orbital decay rates of TrES-5 b to date.

| Reference | Data Sources | Data Points | Period | $\Delta\mathrm{BIC_{LQ}}$ | $\dot{P}$ [ms yr$^{-1}$] |
|---|---|---|---|---|---|
| G. Maciejewski et al. (2021) | Literature, New light curves | 26 | 2009-2020 | 14.9[a] | $-20.4 \pm 4.5$ |
| E. S. Ivshina & J. N. Winn (2022) | Literature, TESS (41)[b] | 73 | 2009-2021 | – | $-17.5 \pm 3.8$ |
| S. R. Hagey et al. (2022) | ETD (DQ 1-3) | 100 | 2012-2021 | 22.2 | $-34.5 \pm 4.6$ |
| W. Wang et al. (2024) | Literature, TESS (41, 55, 56, 57)[b] | 121 | 2009-2022 | 15.4 | $-9.7 \pm 2.1$ |
| L.-C. Yeh et al. (2024) | Literature, TESS (41)[b] | 105 | 2009-2021 | – | $-31.9 \pm 2.2$[c] |

a. A sinusoidal model had similar $\Delta\mathrm{BIC_{LS}} = 13.3$.

b. Respective sectors from TESS.

c. Calculated from value for $dP/dE = -1.5 \times 10^{-9} \pm 1 \times 10^{-10}$ given by L.-C. Yeh et al. (2024).

### 3.1. Data from the Peer-Reviewed Literature

As a starting point for compiling the database for this study, the work of E. S. Ivshina & J. N. Winn (2022) provided both a comprehensive review of the peer-reviewed literature up to 2022 and new fits of light curves from TESS Sector 41. The 73 transit midtimes reported therein were sourced from four distinct studies (G. Maciejewski et al. 2016; E. N. Sokov et al. 2018; G. Maciejewski et al. 2021; E. S. Ivshina & J. N. Winn 2022).

G. Maciejewski et al. (2016) published four new midtimes in BJD$_{\mathrm{TDB}}$ from individual transit observations. Additionally, a total of ten transit midtimes from previous studies (G. Mandushev et al. 2011; D. Mislis et al. 2015) were incorporated, with midtimes provided in BJD$_{\mathrm{TDB}}$. The discovery paper of TrES-5 b, G. Mandushev et al. (2011), did not provide individual transit midtimes, but published a transit ephemeris based on survey data from the Lowell Observatory Planet Search Survey Telescope (PSST; E. W. Dunham et al. 2004, Section 4.1) combined with five subsequent light curves from telescopes at Lowell observatory (originally in HJD$_{\mathrm{UTC}}$). These light curves were provided to G. Maciejewski et al. (2016), who converted the times to BJD$_{\rm TDB}$ and refit them. Because the latter work used a fitting algorithm that assigned a common midtime to all observations at the same transit epoch, only four midtimes were reported in G. Maciejewski et al. (2016), since two transits were observed at different wavelengths during the same night. Similarly, the times initially reported by D. Mislis et al. (2015) were based on a refit of the same five light curves. One midtime published by G. Maciejewski et al. (2016) was credited to D. Mislis et al. (2015) despite not appearing in that work. It was confirmed through private communication that the corresponding light curve had been provided by D. Mislis et al. (2015) along with the other data from that work, yet its midtime got published first in G. Maciejewski et al. (2016). No errors were discovered in the timing conversions, nor were any of the provided uncertainties unrealistically small (e.g. $< 10$ s for a 1-m class telescope). It is noted, however, that composite midtimes (that is, midtimes derived from more than one light curve of the same transit epoch) were often mislabeled as derived from single light curves by E. S. Ivshina & J. N. Winn (2022).

E. N. Sokov et al. (2018) published 30 new individual transit midtimes; most of the corresponding transit light curves were uploaded to the ETD. Those were automatically incorporated into this study if they met certain criteria (Section 3.2).

G. Maciejewski et al. (2021) published twelve new midtimes in BJD$_{\rm TDB}$. On three nights, composite times were reported, which were again not noted as such by E. S. Ivshina & J. N. Winn (2022). As these midtimes showed no signs of errors, they were included in the database.

E. S. Ivshina & J. N. Winn (2022) published 17 individual transit fits from TESS Sector 41 data. These midtimes, provided in BJD$_{\rm TDB}$, were added to the database and were also used to evaluate this work's fitting model (Section 6).

### 3.2. *Data from the Exoplanet Transit Database*

The ETD acted as the second source for this work. All transit midtimes with large uncertainties ($> 5$ min) and poor data quality (using their data quality flag DQ $\geq 3$, indicating lower reliability) were automatically excluded. Three ETD entries were identified as duplicates and thus removed. Timing data, which are provided in HJD$_{\rm UTC}$, were downloaded in bulk and transformed to HJD$_{\rm TDB}$ using the `Astropy` package ( Astropy Collaboration et al. 2013, 2022). This results in the ETD times being in a hybrid HJD$_{\rm TDB}$ format rather than the generally favored BJD$_{\rm TDB}$ timing system, which differ by not more than a few seconds. Given the typical magnitude of the ETD midtime errors, this has minimal impact on the TTV analysis.

The present study does not use the transit midtimes compiled by the Exoclock catalog (A. Kokori et al. 2023). Many transits have been reported to both Exoclock and the ETD, but unfortunately neither site has a complete set of transits: observations are found in each database that are missing from the other. Since transit epochs may have multiple independent observations and there is no standard nomenclature to distinguish between them, untangling which observations are unique and compiling a master database free of duplicates is not trivial and beyond the scope of this work. All but one of the preceding studies used the transits from the ETD; this work does the same to provide more directly comparable results.

### 3.3. *Timing Analysis of Data Through 2022*

As of January 2025, a total of 150 transits were compiled, including 43 from the peer-reviewed literature and 107 from the ETD, covering the period from 2009 to 2024. Two ETD midtimes were found to differ from a best-fit linear ephemeris by more than $10\,\sigma$, indicating a possible timing issue, and discarded. No additional duplicates were found between the data in the peer-reviewed literature and the ETD.

Figure 1 presents an initial TTV analysis intended to replicate prior results using peer-reviewed literature and ETD data up to the end of 2022. The transit timing residuals from the ephemeris models were calculated using linear and quadratic least-squares fits using the `numpy.polyfit` function of the `NumPy` Python library. It shows the residuals from a linear model, also commonly referred to as the observed-minus-calculated (O-C) times. The data set used is approximately equivalent to combining the data available in E. S. Ivshina & J. N. Winn (2022) and S. R. Hagey et al. (2022), except that S. R. Hagey et al. (2022) used 110 points DQ = 1-3 and excluded partial transits, while this work uses 101 points with DQ = 1-2 and does not exclude partials. The quadratic model for this subset of the data has $\dot{P} = -22.9 \pm 5.4\,{\rm ms\,yr^{-1}}$, a $4\,\sigma$ effect, with a corresponding $\Delta{\rm BIC_{LQ}} = 67$, which shows significant support for a decreasing orbital period, and thus a potential for orbital decay. Removing one of the early midtimes (the second point, from G. Mandushev et al. 2011) causes the $\Delta{\rm BIC_{LQ}}$ value to drop by slightly less than 25%, and, taken as a

**Table 2.** Comparison of fits to earliest reported transits.

| Midtime | +1 σ | -1 σ | Diff | Source |
|---|---|---|---|---|
| [BJD_TDB] | [d] | [d] | [s] | |
| 2455152.73184 | 0.00093 | 0.00080 | – | G. Maciejewski et al. (2016) |
| 2455152.73182 | 0.00094 | 0.00071 | -2 | This work |
| 2455352.83535 | 0.00029 | 0.00028 | – | G. Maciejewski et al. (2016) |
| 2455352.83535 | 0.00031 | 0.00027 | 0 | This work |
| 2455444.73500 | 0.00053 | 0.00053 | – | G. Maciejewski et al. (2016)[a] |
| 2455444.73536 | 0.00071 | 0.00053 | +38 | This work, B |
| 2455444.73456 | 0.00065 | 0.00057 | -31 | This work, I |
| 2455530.70451 | 0.00056 | 0.00046 | – | G. Maciejewski et al. (2016) |
| 2455530.70445 | 0.00035 | 0.00031 | -5 | This work |

a. From joint fit to B and I lightcurves.

group, it appears that the five earliest times, all taken from G. Mandushev et al. (2011), have a significant influence on the model and favor a quadratic trend.

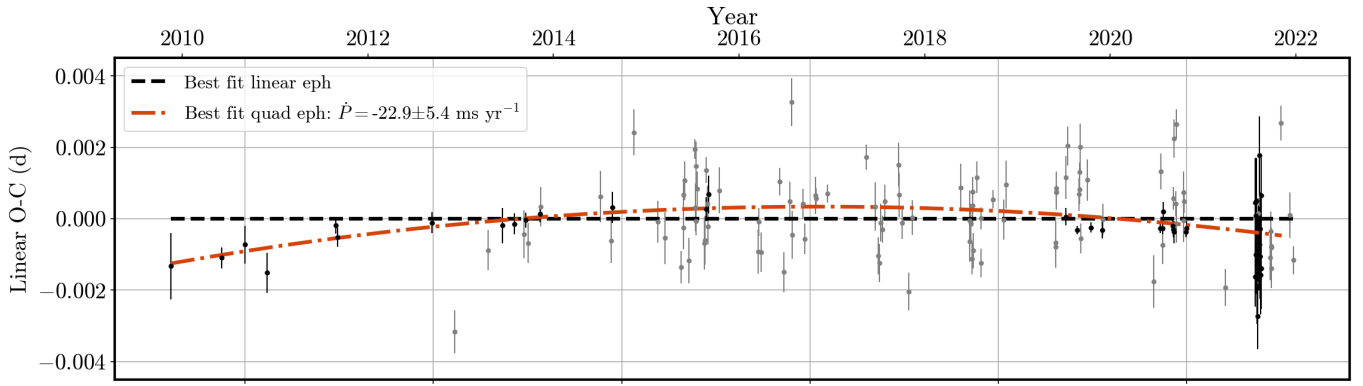

**Figure 1.** Timing residuals of a linear model for data available up to the end of 2022, with the best quadratic fit to this data shown in red. This is equivalent to the data available for the timing analyses of E. S. Ivshina & J. N. Winn (2022) and S. R. Hagey et al. (2022).
Literature midtimes compiled by E. S. Ivshina & J. N. Winn (2022) are shown in black, while those from the ETD are in gray.

To investigate whether there is an issue with the reported midtimes for the earliest transits, five original light curves representing the first four midtimes in the literature record were obtained from the authors of G. Mandushev et al. (2011) and G. Maciejewski et al. (2016), and fitted using `juliet` (N. Espinoza et al. 2019) in the same manner as the other light curves (Section 6). The resulting transit midtimes and errors were very similar (Table 2), and no evidence was found for a possible timing conversion error. Note that two of these transits are partials: (1) The first transit (2009 Nov 17, Lowell 0.8 m) is missing egress, which is reflected in its error bars in both the original and the re-fit midtimes. (2) The fourth transit (2010 Nov 30, Lowell 1.1m) also ends right as egress ends. In addition, two transits observed on the same night (2010 Sep 05, Lowell 1.8 m in B and I filters) appear as a single midtime in G. Mandushev et al. (2011); the value fitted to both transits together in G. Maciejewski et al. (2016) lies precisely between the individual midtimes reported here. Since the midtimes (or average midtimes) of none of these transits disagrees by more than 5 s between the two fitting methods, the original values were used for consistency with past work in the rest of the analysis.

The question therefore remains: is the timing of the five earliest transits evidence of a real departure from a linear ephemeris, or simply a statistical quirk of low number statistics and random noise?

### 3.4. *Non-transit Data on System in Sparse*

Due to a combination of the relative faintness of the target star and competition for observational resources with other exoplanets, the TrES-5 system is relatively unstudied, thus limiting the available data that could be used to complement timing data. Only eight radial velocity measurements spanning 217 days have ever been published (G. Mandushev et al. 2011) and are thus not useful for identifying long-term trends caused by potential planetary or stellar companions. A single high-resolution imaging study to search for close stellar companions was reported by E. N. Sokov et al. (2018). Although no companions were detected down to 1 mag fainter than the host star at separations of 72–1080 au, these modest limits do not exclude a range of plausible configurations involving fainter stars or brown dwarfs.

The rotation period of the star TrES-5 is likewise unknown. No estimate was made in D. Mislis et al. (2015), which examined two UHJs, TrES-5 b and Qatar-1 b, orbiting similar K-type dwarfs, but derived a rotation period for Qatar-1 b only using available long-term photometry. This work examined the two longer sets of photometry available, the PSST data from 2007 and 2009, and TESS data (8 sectors spanning 3 years) using the TESS-SPOC pipeline before flattening. After masking out the transit epochs, a Lomb-Scargle periodogram (N. R. Lomb 1976; J. D. Scargle 1982) was calculated for each TESS sector or PSST segment. No significant or consistent periodicity was uncovered in any of the time series, aside from the transit signal.

Finally, no occultations have been reported for this system. The planet is estimated to be relatively cool for a hot Jupiter ($T_{\rm eq} = 1480 \pm 13$ K; D. Mislis et al. 2015); coupled with the host star's faint magnitude, the detection of an occultation would probably require observations at longer (near-infrared) wavelengths with a moderate- to large-aperture telescope.

## 4. GROUND-BASED OBSERVATIONS

This paper revisits TrES PSST survey data acquired between 2007 and 2009 by G. Mandushev et al. (2011), as the timing of these observations had not yet been independently analyzed.

In addition, fourteen new ground-based transit light curves of TrES-5 b are introduced. These observations were carried out between August 2016 and October 2024 using the 1.23 m Calar Alto Telescope (CA 1.23 m) at Calar Alto Observatory in Spain (one light curve) and the 0.6 m Astronomical Telescope of the University of Stuttgart (ATUS) at Sierra Remote Observatories (SRO) in California (13 light curves). A summary of all observational details is provided in Table 3. Full light curves for each transit, including data from PSST which were not previously published, have been provided online with a stub table for reference in Table 4.

### 4.1. *TrES Survey Data*

Observations of the TrES-5 b field were made by the TrES survey team between UT 2007 July 15 and UT 2007 October 7, and again between UT 2009 June 23 and UT 2009 August 29 on the Lowell Observatory. The original data used to discover the planet were published in G. Mandushev et al. (2011) as a figure and as part of the best transit ephemeris, but no light curve data or individual midtimes were published. Light curve data were kindly provided by G.̃Mandushev and have been fit here independently for the first time; permission was also granted to publish the photometry in this work.

Due to the long cadence ($\approx 500$ s between exposures) and the low signal-to-noise ratio of the PSST data, it was not possible to fit individual transits. Instead, the data were split between the two observing campaigns, and each campaign was phase-folded using the ephemeris published by G. Mandushev et al. (2011) ($T_0 = 2455443.25153$ in HJD$_{\rm UTC}$, $P = 1.4822446$ d). Using this ephemeris, the transit midtime nearest the middle of the campaign was identified as the assigned epoch, and the orbital phase was found for all times in the sequence. Using the corresponding calculated midtime for the assigned epoch, which is set to zero phase, a time was assigned to each data point based on its orbital phase. All times, including the light curve data and the ephemeris, were originally provided in HJD$_{\rm UTC}$ in G. Mandushev et al. (2011), so for this work they were converted to HJD$_{\rm TDB}$ before comparison with other data. The resulting folded transits are shown in Figure 2; the second campaign (P2) had notably better data quality than the first (P1).

### 4.2. *Calar Alto 1.23 m Telescope*

A series of 120 s exposures was taken on the Calar Alto 1.23 m with a DLR-MKIII CCD camera through a Cousins-*R* filter, covering the full transit as well as data before ingress and after egress. The airmass ranged from 1.27 to 2.01.

**Table 3.** Observational details for PSST data and 14 other new ground-based transits.

| # | Date[a] | Telescope | Diameter [m] | Site | Duration [hr] | Binning | $N_{frames}$[b] | Filter | Exposure [s] |
|---|---|---|---|---|---|---|---|---|---|
| P1 | 2007 Jul-Oct | PSST | 0.30 | Lowell | – | – | 1178 | R | 90 |
| P2 | 2009 Jun-Aug | PSST | 0.30 | Lowell | – | – | 1754 | R | 90 |
| 1 | 2016-08-23 | CA 1.23 m | 1.23 | CAO | 2.9 | 1x1 | 507 | R | 120 |
| 2 | 2019-07-23 | ATUS | 0.60 | SRO | 3.1 | 1x1 | 507 | Clear | 20 |
| 3 | 2019-07-29 | ATUS | 0.60 | SRO | 3.1 | 1x1 | 256 | Clear | 35 |
| 4 | 2020-07-29 | ATUS | 0.60 | SRO | 2.8 | 1x1 | 214 | Clear | 45 |
| 5 | 2023-12-13 | ATUS | 0.60 | SRO | 3.2 | 1x1 | 95 | Exo | 120 |
| 6 | 2024-03-18 | ATUS | 0.60 | SRO | 3.2 | 2x2 | 173 | Exo | 70 |
| 7 | 2024-03-21 | ATUS | 0.60 | SRO | 3.8 | 2x2 | 265 | Exo | 50 |
| 8 | 2024-05-03 | ATUS | 0.60 | SRO | 4.0 | 2x2 | 278 | Clear | 50 |
| 9 | 2024-05-09 | ATUS | 0.60 | SRO | 4.7 | 2x2 | 139 | Clear | 120 |
| 10 | 2024-05-12 | ATUS | 0.60 | SRO | 5.6 | 2x2 | 197 | Clear | 100 |
| 11 | 2024-05-21 | ATUS | 0.60 | SRO | 3.8 | 2x2 | 114 | Exo | 120 |
| 12 | 2024-09-15 | ATUS | 0.60 | SRO | 4.2 | 2x2 | 486 | Exo | 30 |
| 13 | 2024-09-30 | ATUS | 0.60 | SRO | 2.8 | 1x1 | 117 | Clear | 80 |
| 14 | 2024-10-03 | ATUS | 0.60 | SRO | 3.6 | 2x2 | 426 | Clear | 10-30 |

a. Date of transit midtime in UT.

b. Number of used exposures.

**Table 4.** Light curves for PSST data and 14 other new ground-based transits[a].

| Transit | Instrument | Time [BJD$_{TDB}$] | Orig. time[b] [HJD$_{UTC}$] | Flux | Flux err |
|---|---|---|---|---|---|
| TrES-5b_20070718_PSST | PSST | 2454339.721130 | 2454299.738781 | 0.99355 | 0.00870 |
| TrES-5b_20070718_PSST | PSST | 2454339.721700 | 2454299.745241 | 0.98780 | 0.01462 |
| TrES-5b_20070718_PSST | PSST | 2454339.722192 | 2454299.752217 | 0.99710 | 0.01023 |
| . . . | | | | | |

a. Light curve photometry for transits in Table 3. This is a stub; full table is available online.

b. Original times for PSST data only (before timing conversion and phase-folding).

The data were reduced using the `defot` aperture photometry pipeline (J. Southworth et al. 2009, 2014) with software apertures of 16, 24, and 50 pixels. No bias or flat-field calibrations were applied, as these have little effect beyond slightly increasing the scatter of the observations. A differential-magnitude light curve was constructed by optimizing the weights of five comparison stars simultaneously with a low-order polynomial fitted to the out-of-transit data. The data were placed on the BJD$_{TDB}$ timescale using routines from J. Eastman et al. (2010). Complete information about this facility is reported in L. Mancini et al. (2017).

### 4.3. *University of Stuttgart's Astronomical Telescope*

ATUS is a 0.6 m, fully reflective f/8 Ritchey–Chrétien telescope mounted on a customized German equatorial mount. It is optimized for photometric measurements of transient events, both in the high-cadence and deep-exposure domain. Its primary instrument is an Andor camera with a back-illuminated, frame-transfer 1k × 1k EMCCD. Each individual image gets time-stamped with an accuracy of better than 1 µs thanks to a GPS-disciplined time reference system which is directly triggered by the camera. Observations were made with either an exoplanet (blue-blocking) bandpass filter or with a clear filter for more throughput. The telescope was located at SRO near Auberry, California until October 2024 and is described in detail in Schindler et al. (in preparation).

Data reduction and differential aperture photometry of all ATUS data were performed using `AstroImageJ` (`AIJ`, K. A. Collins et al. 2017). The reduction process, managed through the integrated `CCD Data Processor`, applied dark and flat-field frames and converted times to $BJD_{TDB}$. Differential aperture photometry was employed using the integrated `Multi-Aperture Tool`. Up to twelve candidate comparison stars were used, each with brightness levels between 50 and 150% of the target star. Aperture radii were updated dynamically throughout each observational window to accommodate varying conditions. Bright background pixels were removed to avoid contamination from stars within the background annulus. A preliminary transit model determined the optimal set of comparison stars by minimizing the BIC. Any frame whose relative flux deviated by more than $5\,\sigma$ from the initial fit was discarded (e.g. aperture contamination through cosmic ray hit or satellite trail). Flux values were detrended against airmass and normalized to an average unity baseline flux.

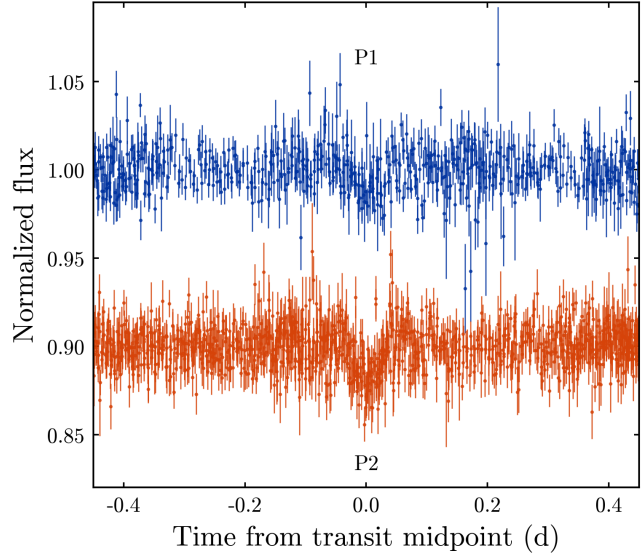

**Figure 2.** Folded transits for PSST data from 2007 (blue) and 2009 (orange), using data taken by G. Mandushev et al. (2011).

## 5. TESS OBSERVATIONS

In addition to new ground-based observations, this study presents a total of 114 new fits to individual light curves obtained from the TESS mission between 2022 and 2024. The analysis encompasses transits from Sectors 55 (17), 56 (19), 57 (16), 75 (19), 77 (11), 82 (17), and 84 (15), utilizing the data with the shortest available cadence for each sector.

The Python package `lightkurve` ( Lightkurve Collaboration et al. 2018) was used to download and analyze the data from the SPOC pipeline; all TESS data utilized in this study can be accessed through the MAST archive ( TESS Team 2021). It was then conditioned to mask segments with excessive noise. Because TESS light curves often exhibit more long-term stellar variability than ground-based data, a detrending step for long-period variability became necessary before fitting. Planetary transit times were masked to interpolate a smoothing function for each sector, using a smoothing window of 501 frames with the built-in "flatten" method. The full time series was divided by the smoothed light curve and normalized. Each sector was then segmented around individual transits.

## 6. LIGHT CURVE FITTING

Transit light curves were analyzed using the publicly available Python package `juliet` (N. Espinoza et al. 2019), which relies on the K. Mandel & E. Agol (2002) analytical approach for transit modeling using `batman` (L. Kreidberg 2015a). The package employs the nested sampler `MultiNest` (F. Feroz et al. 2009, 2019) to efficiently explore the parameter space and provide robust estimates of transit and orbital parameters.

In all fits, joint (Section 6.1) and individual (Section 6.2), the orbital period was fixed to the value taken from the Exoplanet Characterization Catalog, developed for the ExoClock project (A. Kokori et al. 2022), since its precise value has little effect on the other fit parameters. Following G. Maciejewski et al. (2021) a circular orbit was assumed, setting both the orbital eccentricity, *ecc*, and the argument of periastron, *omega*, to zero. This assumption is further discussed in Section 7.1. For the photometric instrument, *mdilution* is the dilution factor,

$$D = \frac{1}{1 + \sum_n \frac{F_n}{F_T}},$$ (8)

with $n$ being the number of blended sources each contributing a flux, $F_n$, and the target star flux, $F_T$, all located inside the aperture. Since bright background pixels were removed in `AIJ` when performing the aperture photometry, this value was set to 1. The quadratic limb darkening coefficients, $u_1$ and $u_2$, were calculated in `juliet` using PHOENIX models (A. Claret & S. Bloemen 2011), together with the filter response curves and the stellar parameters. `juliet` does not incorporate these coefficients directly, but instead uses the correlations proposed by (D. M. Kipping 2012):

$$q_1 = (u_1 + u_2)^2,$$ (9a)

$$q_2 = \frac{u_1}{2(u_1 + u_2)}.$$ (9b)

By default, two instrumental parameters remained free. The parameter *mflux* represents the offset in relative flux for the photometric instrument, while *sigma*$_w$ captures a jitter term added in quadrature to the error bars (in parts per million) to account for any noise not otherwise accounted for. Best-fitting values emerged from the medians of the marginalized posterior distributions, with $1\,\sigma$ intervals determined by the $15.9\,\%$ and $84.1\,\%$ percentiles.

### 6.1. Joint Fit to Determine System Parameters

To determine the best system parameters, a joint fit model with the 14 new ground-based light curves was calculated, excluding the TrES survey data. The fitted parameters include the ratio of orbital semi-major axis to stellar radius, $a$, the ratio of planet to stellar radius, $p$, and the impact parameter, $b$, and were constrained to take on the same value for all light curves. A transit midtime, $t_0$, was fit for each light curve. Following N. Espinoza (2018), the parameters $r_1$ and $r_2$ served to efficiently sample $p$ and $b$. The coefficients $q_1$ and $q_2$ were held fixed and linked for light curves acquired with the same filter. To ensure thorough exploration, the fit was performed with 700 live points, which are the active samples maintained by the nested sampling algorithm to explore the parameter space and compute the Bayesian evidence. Convergence was verified by inspecting the posterior distributions and parameter correlations. The refined transit parameters are listed in Table 5, including parameters computed from the model outputs, with uncertainties propagated accordingly. The results are consistent within $1\,\sigma$ of previous studies, demonstrating the precision achievable with well-maintained small- to mid-sized ground-based telescopes. The best-fitting transit model for each light curve is presented in Figure 3.

**Table 5.** System parameters derived from a joint-fit transit model of 14 new ground-based observations.

| Parameter | This Work |
| --- | --- |
| Semi-Major Axis to Stellar Radius | $6.1465^{+}0.093_{-}0.069$ |
| Planet Radius to Stellar Radius | $0.14097^{+}0.00063_{-}0.00074$ |
| Transit Depth [ppth] | $19.872^{+}0.178_{-}0.208$ |
| Impact Parameter | $0.573^{+}0.014_{-}0.019$ |
| Inclination [°] | $84.65^{+}0.26_{-}0.19$ |
| Transit Duration [min] | $110.7^{+}1.8_{-}1.5$ |
| Stellar Density [$\rho_\odot$] | $1.420^{+}0.064_{-}0.048$ |

### 6.2. Individual Fits to Determine Midtimes

The set of priors for the individual fits are summarized in Table 6. Most parameters were fixed to the best-fit values obtained from the joint fit. Each light curve, now including the TESS and PSST survey data, was fit individually to determine the transit midtime $t_0$. For the individual fits, 500 live points were used.

For each fit, the reduced $\chi^2$ value was calculated to determine the goodness of fit for the transit model. For the 13 ATUS transits, a broad range of $\chi^2$ values from 1.1 to 3.4 was found, with most light curves well fit by $\chi^2 < 1.5$, though six curves showed $1.7 < \chi^2 < 3.4$. The point-to-point noise assigned by AIJ to the transit flux varied considerably from transit to transit; in particular the earliest three transits from 2019-2020 had smaller assigned errors and also the worst $\chi^2$ values ($\chi^2 > 3$). Lower errors may have been due to including variable observing strategies (exposure times ranged from 10-120 s and both open and exoplanet filters were used); telescope cleanings and upgrades; and variable weather conditions. This is not the case for the light curves obtained from PSST, CAHA, or TESS, all of which had reduced $\chi^2 < 1.4$, indicating reasonable assigned errors. Thus, for the ATUS light curves only the nominal flux errors were rescaled by a factor of $\chi$ based on the initial fit, which increased the midtime errors of these transits by factors ranging from 1.04 to 1.84. Midtimes for the new fits to ground-based transit are listed in Table 7. All midtimes used including those from TESS, the ETD, and the peer-reviewed literature, are included in the online version of this table.

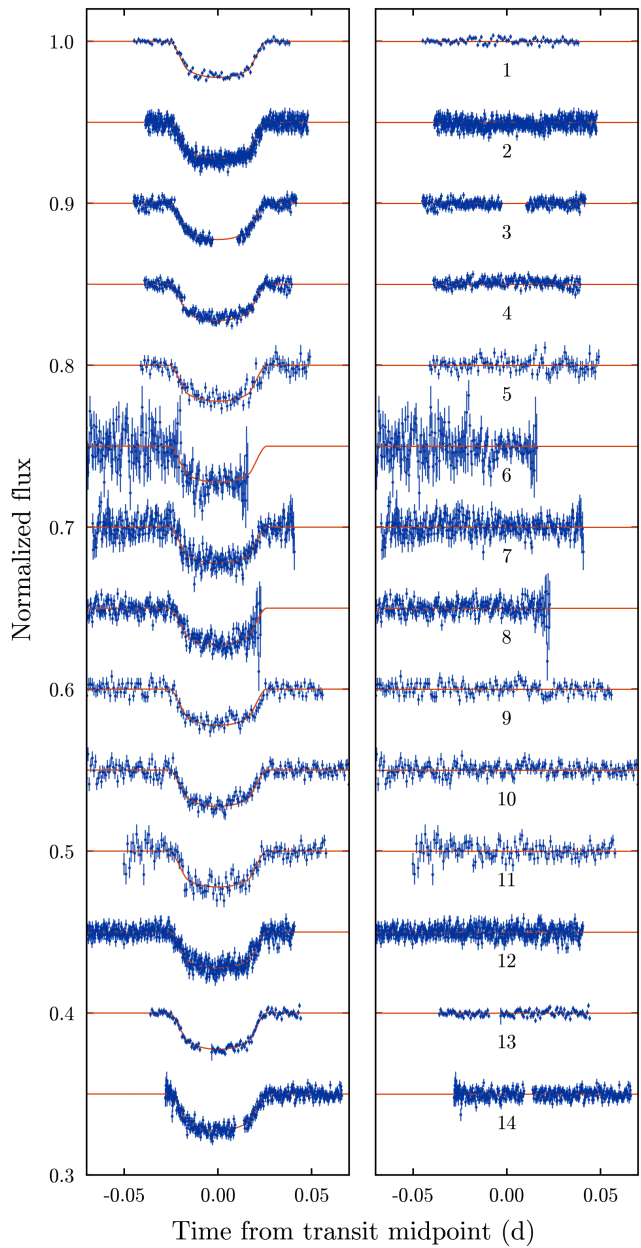

**Figure 3.** *Left:* Joint-fit transit model of 14 new ground-based observations, as numbered in Table 3. *Right:* Residuals from the model fit to each light curve. Transits and models were normalized to 1 for out of transit flux, and then offset by 0.05 for display.

**Table 6.** Initial planetary and instrumental parameters and their prior distributions in `juliet`.

| Parameter | Distribution | Hyperparameter |
|---|---|---|
| $P$ | fixed | 1.482246865 |
| $t_0$ | normal | $[t_{ref}^a, 0.1]$ |
| $a$ | fixed | 6.15 |
| $e$ | fixed | 0 |
| $\omega$ | fixed | 0 |
| $p$ | fixed | 0.141 |
| $b$ | fixed | 0.573 |
| $mdilution$ | fixed | 1.0 |
| $mflux$ | normal | $[0., 0.1]$ |
| $\sigma_w$ | loguniform | $[0.1, 1000]$ |
| $q_1$ | fixed | $q_1^b$ |
| $q_2$ | fixed | $q_2^b$ |

a. Initial guess using reference epoch (2457011.470352) and orbital period, $P$.
b. Depending on the chosen filter: R ($q_1 = 0.44923$, $q_2 = 0.41054$), Clear/Open ($q_1 = 0.48374$, $q_2 = 0.44156$), Exo ($q_1 = 0.42875$, $q_2 = 0.41203$), and TESS ($q_1 = 0.34300$, $q_2 = 0.38708$).

**Table 7.** New ground-based midtime values added by this study.

| # | Telescope[a] | Time system | Midtime | $+1\,\sigma$ | $-1\,\sigma$ |
|---|---|---|---|---|---|
| | | | [d] | [d] | [d] |
| P1 | PSST | $HJD_{TDB}$ | 2454340.46624 | 0.00144 | 0.00152 |
| P2 | PSST | $HJD_{TDB}$ | 2455044.52935 | 0.00113 | 0.00108 |
| 1 | CA 1.23 m | $BJD_{TDB}$ | 2457623.63768 | 0.00025 | 0.00028 |
| 2 | ATUS | $BJD_{TDB}$ | 2458687.89053 | 0.00020 | 0.00020 |
| 3 | ATUS | $BJD_{TDB}$ | 2458693.81929 | 0.00019 | 0.00019 |
| 4 | ATUS | $BJD_{TDB}$ | 2459059.93454 | 0.00025 | 0.00023 |
| 5 | ATUS | $BJD_{TDB}$ | 2460291.68197 | 0.00066 | 0.00067 |
| 6 | ATUS | $BJD_{TDB}$ | 2460388.02912 | 0.00155 | 0.00134 |
| 7 | ATUS | $BJD_{TDB}$ | 2460390.99228 | 0.00045 | 0.00049 |
| 8 | ATUS | $BJD_{TDB}$ | 2460433.97847 | 0.00058 | 0.00061 |
| 9 | ATUS | $BJD_{TDB}$ | 2460439.90679 | 0.00056 | 0.00058 |
| 10 | ATUS | $BJD_{TDB}$ | 2460442.87085 | 0.00056 | 0.00051 |
| 11 | ATUS | $BJD_{TDB}$ | 2460451.76453 | 0.00069 | 0.00070 |
| 12 | ATUS | $BJD_{TDB}$ | 2460568.86187 | 0.00031 | 0.00029 |
| 13 | ATUS | $BJD_{TDB}$ | 2460583.68377 | 0.00023 | 0.00024 |
| 14 | ATUS | $BJD_{TDB}$ | 2460586.64865 | 0.00021 | 0.00024 |

a. Online table shows the full database including peer-reviewed literature, ETD and TESS midtimes for Sectors 55 (17), 56 (19), 57 (16), 75 (19), 77 (11), 82 (17), and 84 (15).

### 6.3. *Validation of Fitting Method*

The reliability of this work's fitting approach was tested by comparing fits of midtimes and errors from TESS Sector 41 to those published by E. S. Ivshina & J. N. Winn (2022). That work used a custom Markov Chain Monte Carlo (MCMC) method to calculate the midtimes and their uncertainties, but otherwise similar methodology and values, with the ratio of planet to stellar radius, the ratio of orbital semi-major axis to stellar radius, the impact parameter, and the limb-darkening parameters all fixed. Their free parameters were the transit midtime and the coefficients of a detrending polynomial.

On average, the transit midtimes are in excellent agreement. The times reported by E. S. Ivshina & J. N. Winn (2022) have a median difference of 13 s compared to those calculated using the present study's approach, insignificant compared to the large error bars (median error: 95 s) of TESS data points. Notably, though, midtime values determined here using `juliet` with `MultiNest` have consistently smaller estimated uncertainties, by roughly 15 % across various fits compared to other fitting methods, e.g. using the package `pylightcurve` (A. Tsiaras et al. 2016) as in E. R. Adams et al. (2024). The cause of this discrepancy in errors is worth investigating in the future.

## 7. UPDATED TIMING ANALYSIS

As discussed in Section 3.3, the initial timing analysis examined the literature data from 2009-2022, spanning roughly 13 years or 3000 epochs. This work has added a substantial number of new midtimes (Table 7). The time span of observations has increased by about a third, or four years, now covering 2007-2024 (17 years or 4000 epochs), and the number of available data points nearly doubled to 280.

### 7.1. Periodic Model

Periodic signals may result from orbital precession driven by tidal interactions, from gravitational perturbations from additional celestial bodies, or from stellar line-of-sight acceleration (as hypothesized for WASP-4 b, L. G. Bouma et al. 2020). The first search for periodic signals in TrES-5 b was conducted by E. N. Sokov et al. (2018), who identified a signal with a period of around 99 days. G. Maciejewski et al. (2021) also detected this approximately 99-day periodicity but found it was not statistically significant, as all tested periods shorter than roughly 3000 days exhibited false alarm probabilities (FAPs) exceeding 5%. Instead, a long-term trend with a period of 31,000 days was proposed as the most promising scenario, favored by a significant $\Delta BIC_{LS}$ value of 13.3 over the linear model.

To investigate whether any sinusoidal component could be robustly extracted from the updated timing dataset, the linear model was first subtracted from the transit midtime data to obtain the timing residuals. Since transit midtime data are sampled no more often than the orbital period of the planet, the minimum period that can be searched for is the Nyquist period $P_N = 2 \times P = 2.96$ days. A Lomb-Scargle periodogram (N. R. Lomb 1976; J. D. Scargle 1982) was calculated on the full set of these residuals searching for periods between 3 and 100,000 days in order to detect a putative 31,000-day period, as shown in the top panel of Figure 4. However, the effective maximum period that can be meaningfully probed is half the time spanned by the dataset, or about 3,000 days, beyond which the Lomb-Scargle algorithm does not return any peaks.

The top-three locally significant peaks were then calculated, with an arbitrary distance of 30 days between peaks to avoid clustering all of the detections at a single period and its near aliases. Only one peak was found with a FAP of less than than 1% (at 5.2 days); the next closest peaks (at 85 and 2736 days) have much higher FAPs ($> 5\%$ and 95%, respectively). However, it is important to consider that these FAP values should only serve as general indicators, as their calculation inherently assumes the uncertainties in the data points are independent and identically distributed random variables (J. T. VanderPlas 2018). This assumption likely does not hold for heterogeneous datasets containing errors and measurements derived from diverse sources.

Using each of these three values as an initial guess, the best sinusoid was then fit to the timing data using Equation 4. In all cases, the fitted amplitudes are small ($P = 5.2$ days: $A = 7.3 \pm 5.8$ s; $P = 85.9$ days: $A = 18.1 \pm 6.0$ s; $P = 4175$ days: $A = 22.5 \pm 6.6$ s) and at low significance (1.3, 3.0, and 3.4 $\sigma$, respectively). Note that the longer period is not well constrained by the available data, hence the jump from an initial guess of $\sim 3000$ days to the fitted $\sim 4000$ days.

Motivated by the number of sub-significant peaks in the full dataset, a second search for periodicities was performed using only the eight TESS sectors, as seen in the bottom panel of Figure 4. For consistency, the midtimes for Sector 41 were conducted using the same method as for the other seven sectors, rather than adopting the values from E. S. Ivshina & J. N. Winn (2022). Although the TESS data spans just 1/6 of the time of the full dataset (2021-2024), it has the advantage of regular sampling, with an average distance between data points of one transit epoch (in the full dataset, there has been on average only one observation every 15 transit epochs). Moreover, the TESS data spanned nearly 1,200 days, sufficient to test whether the 5.2 and 85.9-day periods still appear. However, no significant peaks were detected; moreover, none of the three highest peaks corresponds to a period identified from the full dataset.

As a final point of analysis, the marginal signal at $\sim 2700$ days (not ruled out in the TESS data since it lacks the baseline to detect it) was analyzed using a $\Delta BIC_{LS}$ evolution model similar to that in Figure 1. Although the $\Delta\mathrm{BIC_{LS}}$ value for the sinusoidal fit peaked around 2019 with a value of 60, it has declined notably in recent years and is currently below 20. There are also many outliers, indicating that the values derived for this fit are highly sensitive to individual points in the dataset.

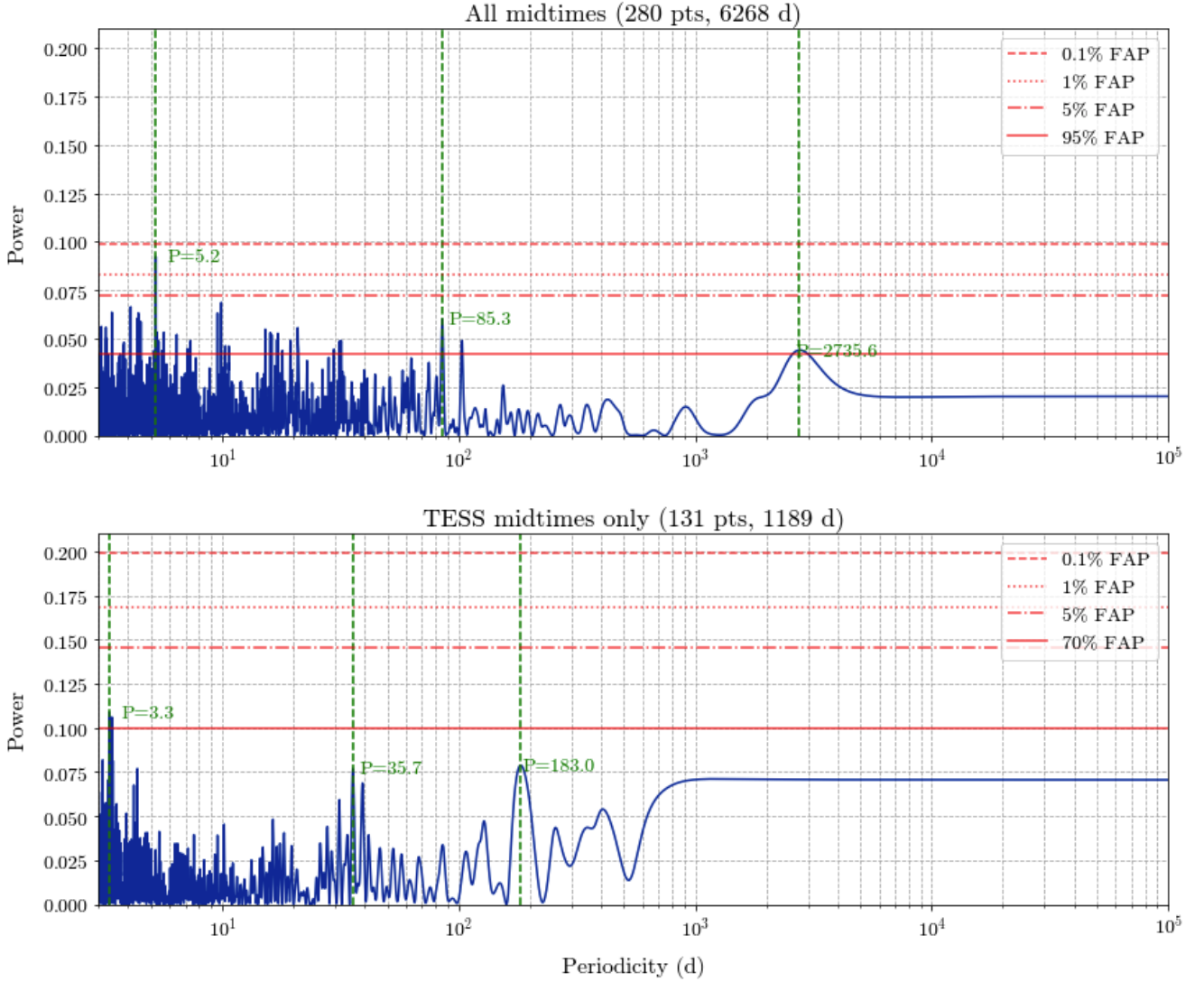

**Figure 4.** Lomb-Scargle periodograms (N. R. Lomb 1976; J. D. Scargle 1982) for all 280 midtimes used in this work (*top*), and for a subset of midtimes derived solely from TESS observations (*bottom*). The full dataset exhibits much higher levels of noise, with one signal (at 5.2 days) above a nominal 0.1% FAP. This peak, however, has only a marginal ($2\,\sigma$) detection in amplitude, and is not seen using just the TESS data. Green-dashed vertical lines highlight locally significant peaks, none of which are consistent between the full dataset and the TESS subset. The TESS data also do not evince any significant periods with FAP less than 70%.

Combining all the evidence, there is no case for a sinusoidal signal anywhere between 3-3000 days in these data, and no timing data available to search for periodicities outside that range.

### 7.2. *Quadratic Model*

In Figure 1, a quadratic model was favored over a linear model by $\Delta\mathrm{BIC_{LQ}} = 67$, with an estimated period decrease rate of $\dot{P}_{\mathrm{init}} = -22.9 \pm 5.4$ ms yr$^{-1}$.

An updated TTV analysis is shown in Figure 5. The inferred orbital decay rate has substantially decreased to $\dot{P} = -5.3 \pm 2.2$ ms yr$^{-1}$ (top panel), less than half of the initial estimate. With the inclusion of new data, the previously rising trend in $\Delta\mathrm{BIC_{LQ}}$ has reversed, lowering the $\Delta\mathrm{BIC_{LQ}}$ from 67 to 11, thereby weakening the case for a quadratic model.

**Table 8.** Best ephemeris fit parameters based on 280 midtimes of TrES-5 b.

|  | Linear (preferred) |
| --- | --- |
| $T_0$ [BJD$_{\rm TDB}$] | 2459443.836676(57) |
| $P$ [d] | 1.482246448(57) |
|  | Quadratic |
| $\Delta\mathrm{BIC}_{LQ}$ | $11.1 \pm 20$ |
| $T_0$ [BJD$_{\rm TDB}$] | 2459443.836691(57) |
| $P$ [d] | 1.48224625(10) |
| $\dot{P}$ [ms yr$^{-1}$] | $-5.3 \pm 2.2$ |
| Corresponding $Q'_\star$ | $5 \times 10^4$ |

The sharp, early rise of the actual data (red curve, middle panel) in $\Delta\mathrm{BIC}_{LQ}$ compared to the quasi-analytic estimate calculated following B. Jackson et al. (2023) (blue curve), which tracks the path an idealized $\Delta\mathrm{BIC}$ evolution would take with observations at the epochs observed, strongly suggests that statistical outliers may have artificially strengthened the preference for a quadratic model in the past. Most critically, the bottom panel shows that the very first data points of G. Mandushev et al. (2011) still have a particularly strong impact on the fit of the quadratic model. Removing just one of the early data points would cause the $\Delta\mathrm{BIC}_{LQ}$ value to drop significantly.

It is noted that in neither model does the reduced $\chi^2$ value reach 1. The best O-C fit with a linear ephemeris has a reduced $\chi^2 = 3.0$ while the quadratic fit has reduced $\chi^2 = 2.9$. A likely explanation for at least some of this discrepancy is underestimated errors in some or all of the transit midtimes. Following E. R. Adams et al. (2024), a rescaling test was performed wherein the errors on all midtimes were increased by a factor of $\chi = 1.7$. In this experiment, the shape of the evolution of $\Delta\mathrm{BIC}_{LQ}$ is unchanged, but the magnitude of $\Delta\mathrm{BIC}_{LQ}$ decreases, so that $\Delta\mathrm{BIC}_{LQ}$ peaks at 12 and has a final value of 0.1, indicating no statistical preference for either model. This shows that underestimated errors could explain all of the $\Delta\mathrm{BIC}_{LQ}$ from the nominal fit.

If this significantly lower value of $\dot{P} = -5.3 \pm 2.2$ ms yr$^{-1}$ still reflected orbital decay due to tidal dissipation, it can be used to estimate the modified tidal quality parameter of the host star (P. Goldreich & S. Soter 1966),

$$Q'_\star = -\frac{27\pi}{2}\left(\frac{M}{M_\star}\right)\left(\frac{a}{R_\star}\right)^{-5}\left(\frac{\mathrm{d}P}{\mathrm{d}E}\right)^{-1} P, \tag{10}$$

with $\frac{M}{M_\star}$ being the ratio of planetary to stellar mass (values taken from A. S. Bonomo et al. 2017) and $\frac{a}{R_\star}$ being the ratio of orbital semi-major axis to stellar radius ($a$ in `juliet`). The best-fit decay rate corresponds to $Q'_\star = 5 \times 10^4$, which, while higher than the previous estimate of $Q'_\star = 1.4 \times 10^4$ found by G. Maciejewski et al. (2021), is still lower than both theoretical predictions ($\approx 10^6$, N. Weinberg, private communication and N. N. Weinberg et al. 2024) and value derived from the observed decay of WASP-12 b ($Q'_\star = 1.75 \times 10^5$, S. W. Yee et al. 2020). Furthermore, while WASP-12 may be a possible subgiant (see discussion in P. Leonardi et al. 2024), TrES-5 is unlikely to also be one.

Combining all of the evidence available – the unusually low value of $Q'_\star$, the non-significant detection ($2.4\,\sigma$) of $\dot{P}$, the much more muted preference for a quadratic model that has rapidly diminished in significance as measured by $\Delta\mathrm{BIC}_{LQ}$ with longer observational baselines, and the fact that the model preference is significantly depend on singly data points – leads to the conclusion that there is no evidence for orbital decay in TrES-5 b.

## 8. DISCUSSION AND CONCLUSIONS

In total, 280 midtimes were used to derive an updated ephemeris for TrES-5 b. Table 8 shows that a linear model provides the best fit, yielding a refined ephemeris of $T_0 = 2459443.836676(57)$ BJD$_{\rm TDB}$ and $P = 1.482246448(57)$ d.

The TrES-5 b timing data have exhibited putative signs of non-linearity since 2018 (E. N. Sokov et al. 2018), with at least five subsequent studies reporting modestly significant values indicating a decreasing orbital period (G. Maciejewski et al. 2021; E. S. Ivshina & J. N. Winn 2022; S. R. Hagey et al. 2022; W. Wang et al. 2024; L.-C. Yeh et al. 2024). The values for the decay rates presented in Table 1 vary by a factor of four, though interpretation is complicated since each value is derived from different subsets of the available transit midtimes. Examining the three most directly comparable datasets, which sequentially incorporate literature data along with an increasing number of TESS sectors, there was a clear trend of decreasing $\dot{P}$ values even prior to this analysis. Specifically, G. Maciejewski et al. (2021) derived $\dot{P} = -20.4 \pm 4.5$ ms yr$^{-1}$ using all data available through 2020. Adding one TESS sector and extending the

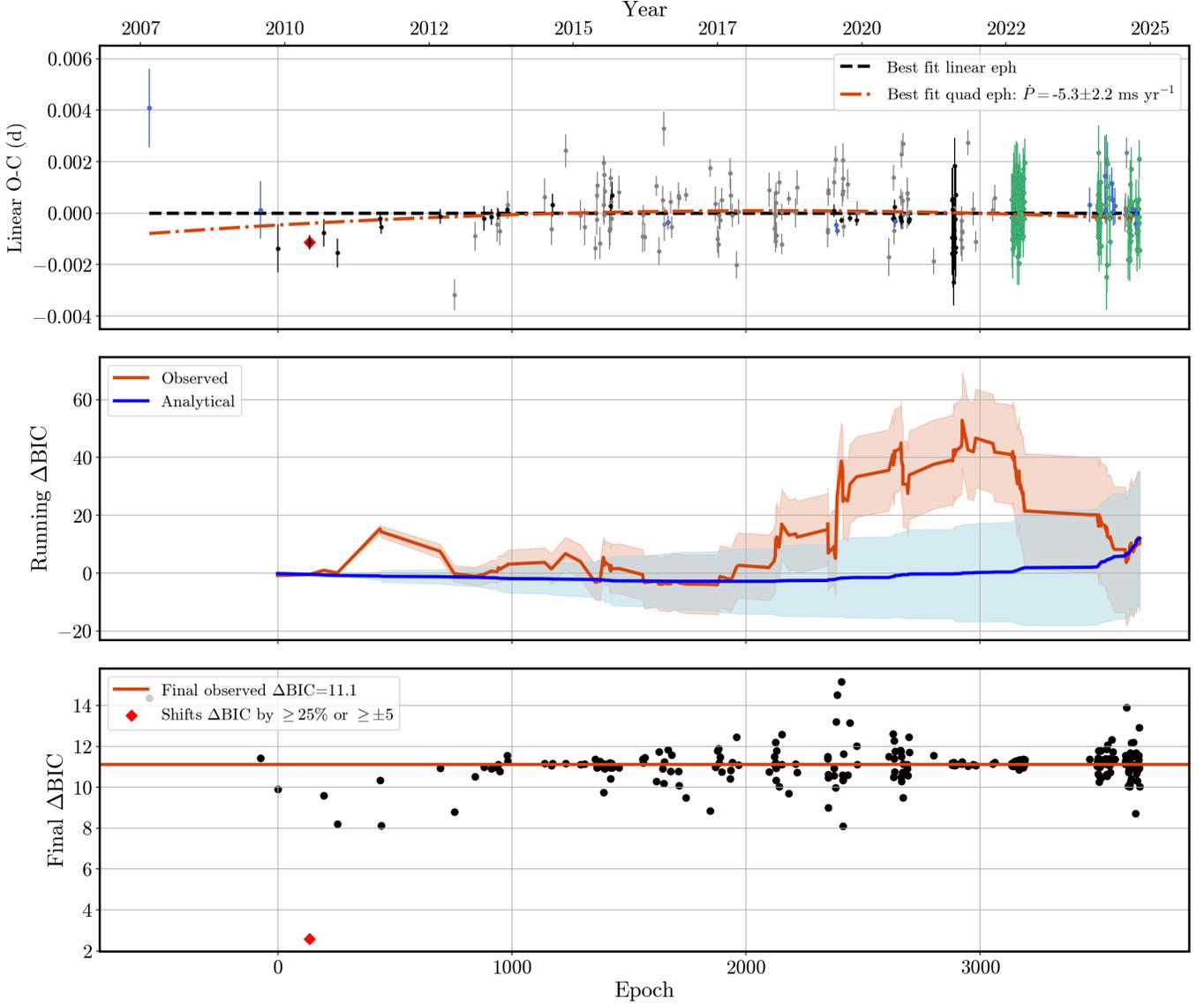

**Figure 5.** Timing residuals of a linear vs. a quadratic model after adding new data (2007-2024). *Top:* Observed-minus-calculated midtimes assuming a linear ephemeris. New midtimes derived from TESS light curves are shown in green, while new ground-based observations are in blue. Literature midtimes compiled by E. S. Ivshina & J. N. Winn (2022) are in black, and those from the ETD are in gray. *Middle:* Evolution of $\Delta\mathrm{BIC_{LQ}}$. The red curve represents the value calculated using Equation 6, while the blue curve depicts the quasi-analytic estimate of B. Jackson et al. (2023), which is the trajectory $\Delta\mathrm{BIC}$ would be expected to take with observations at these epochs if the best-fit $\dot{P}$ corresponded to real orbital decay. Both are accompanied by shaded regions indicating their expected $1\,\sigma$ variations (Equation 7). The calculation begins only after three data points are available, reflecting the parameter fitting requirements of the quadratic model. *Bottom:* Final observed $\Delta\mathrm{BIC_{LQ}}$ (red line, calculated using all data points, see right end of red curve in middle panel), compared to the final observed $\Delta\mathrm{BIC_{LQ}}$ that results after omitting each individual data point.

observations through 2021, E. S. Ivshina & J. N. Winn (2022) reported a slightly reduced rate of $\dot{P} = -17.5 \pm 3.8$ ms yr$^{-1}$. With four TESS sectors and observations through 2022, W. Wang et al. (2024) found the decay rate had halved relative to two years earlier, reporting $\dot{P} = -9.7 \pm 2.1$ ms yr$^{-1}$. Incorporating two recent years of data (including four new TESS sectors) and two earlier years with larger uncertainties, this current analysis finds the decay rate has halved once more to $\dot{P} = -5.3 \pm 2.2$ ms yr$^{-1}$.

The corresponding values for $\Delta\mathrm{BIC_{LQ}}$ also appear to have peaked around 2021 (Figure 5); using this uniform analysis for values of $\Delta\mathrm{BIC_{LQ}}$, the peak was around 50, and has now dropped to 11. Even more critically, the value significantly drops if a single early transit with low error bars is removed. The $1\,\sigma$ errors on $\Delta\mathrm{BIC_{LQ}}$ also place it firmly consistent with zero, with a range of $\pm 20$.

Both the high dependence on a small number of points and the decreasing trend in $\Delta\mathrm{BIC_{LQ}}$ overall suggest that the initially observed trend was a statistical quirk resulting mainly from the low error bars on the second transit from G. Mandushev et al. (2011). There is also an element of low-number statistics in play. The amount by which the early transits deviate from expectations is similar to that of later observations. However, having 10-20 transits per year at later epochs compared to 2-3 at the earliest epochs means that in recent years no single outlying point has a strong impact on observed $\Delta\mathrm{BIC_{LQ}}$. In contrast with other systems (e.g., OGLE-TR-111b, WASP-19b, CoRoT-2b) where early detections of apparent transit timing deviations were due to errors in earlier analyses (E. R. Adams et al. 2010, 2024), the case for TrES-5b does not seem to result from mistakes but rather from an initial signal that disappears with further data.

Additionally, although there is still fairly high scatter in the transit midtime residuals, there is no evidence for either precession or for line of sight acceleration, as postulated by G. Maciejewski et al. (2021) in the absence of other viable explanations for their apparent trend. No periodicity was detected at either shorter time scales ($\sim$ tens to hundreds of days), which would have completed several cycles during the available time series, or at longer timescales ($\sim 3000$ d) comparable to the full data span (see Figure 4).

Finally, it is instructive to note the effect of a single, concerted observing campaign, of the type available to dedicated small-aperture facilities like ATUS, in comparison to TESS observations in a single season. With each additional TESS sector, the value for $\Delta\mathrm{BIC_{LQ}}$ in this system fell; but a similar effect was also seen if the TESS data were excluded and only the 14 ground-based observations were included. Although most bright UHJ systems are regularly observed with TESS, about 13% of UHJs fall through observing gaps with either 0 or 1 TESS sectors available, and another 8% were skipped in the most recent TESS cycle for their hemisphere, which can mean a 4+ year gap in data. Small telescopes are well poised to observe during these gaps in TESS observations. Importantly, the timing precision of ground-based observations—especially from dedicated facilities—can be significantly better than that of TESS. In this case, the median uncertainty of the ground-based midtimes was smaller than for the TESS-derived midtimes. Thus, even a few ground-based observations per year using a 0.6 m telescope may be just as constraining as a full TESS season. The chief advantage of regular observations is to avoid situations where only a few data points are available (as at the beginning of the TrES-5 b time sequence), wherein slight random fluctuations may give the appearance of a detection of orbital decay where none actually exists.

## ACKNOWLEDGMENTS

This study was supported by grants from NASA's Exoplanet Research Program 80NSSC22K0317 and 80NSSC25K7170.

Thanks to G. Mandushev and G. Maciejewski for kindly both providing data and answering questions about data originally taken over 15 years ago.

ATUS was made possible through funding from the Deutsches Zentrum für Luft- und Raumfahrt e.V. (DLR; German Aerospace Center) and the University of Stuttgart in support of *SOFIA*, the *Stratospheric Observatory for Infrared Astronomy*, as well as quality assurance funds allocated by the State of Baden-Württemberg for improving the quality of teaching and learning.

This paper includes data collected with the TESS mission, obtained from the MAST data archive at the Space Telescope Science Institute (STScI). Funding for the TESS mission is provided by the NASA Explorer Program. STScI is operated by the Association of Universities for Research in Astronomy, Inc., under NASA contract NAS 5–26555. This research also made use of the Astrophysics Data System, funded by NASA under Cooperative Agreement 80NSSC21M00561.

LM acknowledges the financial contribution from PRIN MUR 2022 project 2022J4H55R.

*Facilities:* ATUS, CAO:1.2m - Calar Alto Observatory's 1.2 meter Telescope, TESS

*Software:* Astropy ( Astropy Collaboration et al. 2022, 2018, 2013), Matplotlib (J. D. Hunter 2007), pylightcurve (A. Tsiaras et al. 2016), Numpy (C. R. Harris et al. 2020), SciPy (P. Virtanen et al. 2020), emcee (D. Foreman-Mackey et al. 2013a,b), ExoTETHyS (G. Morello et al. 2020), AstroImageJ (K. Collins & J. Kielkopf 2013), Juliet (N. Espinoza et al. 2018), batman (L. Kreidberg 2015b), MultiNest (F. Feroz et al. 2011), PyMultiNest (J. Buchner 2016; J. Buchner et al. 2014), lightkurve ( Lightkurve Collaboration et al. 2018), defot (J. Southworth et al. 2009, 2014)